\documentclass[11pt]{article}
\usepackage[a4paper]{geometry}
\usepackage[english]{babel}

\usepackage{amsmath, accents}
\usepackage{amsfonts}
\usepackage{amstext}
\usepackage{amssymb}
\usepackage{amsthm}
\usepackage{amscd}
\usepackage{mathrsfs}

\usepackage{fancyhdr}
\usepackage[pagebackref,draft=false]{hyperref}
\hypersetup{colorlinks,
linkcolor=myrefcolor,
citecolor=mycitecolor,
urlcolor=myurlcolor}

\usepackage[capitalize]{cleveref}
\usepackage{cite}
\usepackage{caption}
\usepackage{etaremune}

\usepackage{xcolor}
\definecolor{myurlcolor}{rgb}{0,0,0.4}
\definecolor{mycitecolor}{rgb}{0,0.5,0}
\definecolor{myrefcolor}{rgb}{0.5,0,0}
\usepackage{graphicx}
\usepackage{tikz}
\usepackage{tikz-cd}
 \usetikzlibrary{decorations.pathmorphing}

\usepackage{etoolbox}
\usepackage{makeidx}
\usepackage{sectsty}
\usepackage{enumitem} 
\usepackage[]{latexsym}
\usepackage{braket}
\usepackage{caption}
\usepackage[utf8]{inputenx}
\usepackage[T1]{fontenc}
\usepackage{lmodern}
\usepackage{textcomp}
\usepackage{microtype}
\usepackage{totcount}
\usepackage{blindtext}

\newtheorem{example}{Example}

\newtheorem*{proof*}{Proof}

\newcommand{\be}{\begin{equation}}
\newcommand{\ee}{\end{equation}}
\newcommand{\bea}{\begin{eqnarray}}
\newcommand{\eea}{\end{eqnarray}}

\newcommand{\grit}[1]{{\bfseries {\itshape {#1}}}}

\newcommand{\ra}{\rightarrow}

\newcommand{\lra}{\longrightarrow}

\newcommand{\hh}{\mathcal{H}}

\newcommand{\lag}{\mathfrak{L}}

\newcommand{\dd}{{\rm d}}

\title{Covariant Variational Evolution and Jacobi Brackets: Particles}
\date{}

\author{F. M. Ciaglia$^{1,7}$ \href{https://orcid.org/0000-0002-8987-1181}{\includegraphics[scale=0.7]{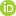}}, F. Di Cosmo$^{2,3,8}$ \href{https://orcid.org/0000-0003-0256-5913}{\includegraphics[scale=0.7]{ORCID.png}}, A. Ibort$^{2,3,9}$ \href{https://orcid.org/0000-0002-0580-5858}{\includegraphics[scale=0.7]{ORCID.png}}, \\ G. Marmo$^{4,5,10}$ \href{https://orcid.org/0000-0003-2662-2193}{\includegraphics[scale=0.7]{ORCID.png}}, L. Schiavone$^{3,4,6,11}$  \href{https://orcid.org/0000-0002-1817-5752}{\includegraphics[scale=0.7]{ORCID.png}} \\
\footnotesize{$^{1}$\textit{ Max Planck Institute for Mathematics in the Sciences, Leipzig, Germany}} \\
\footnotesize{$^{2}$\textit{ ICMAT, Instituto de Ciencias Matem\'{a}ticas (CSIC-UAM-UC3M-UCM)}} \\
\footnotesize{$^{3}$\textit{Depto. de Matem\'aticas, Univ. Carlos III de Madrid, Legan\'es, Madrid, Spain}} \\
\footnotesize{$^{4}$\textit{ INFN-Sezione di Napoli, Naples, Italy}} \\
\footnotesize{$^{5}$\textit{ Dipartimento di Fisica ``E. Pancini'', Universit\`a di Napoli Federico II,  Naples, Italy}} \\
\footnotesize{$^{6}$\textit{ Dipartimento di Matematica e Applicazioni "Renato Caccioppoli", Università di Napoli Federico II, Napoli, Italy}} \\
\footnotesize{$^{7}$\textit{ e-mail: \texttt{florio.m.ciaglia[at]gmail.com} and \texttt{ciaglia[at]mis.mpg.de}}} \\
\footnotesize{$^{8}$\textit{ e-mail: \texttt{fcosmo[at]math.uc3m.es}}} \\
\footnotesize{$^{9}$\textit{ e-mail: \texttt{albertoi[at]math.uc3m.es}}} \\
\footnotesize{$^{10}$\textit{ e-mail: \texttt{marmo[at]na.infn.it}}} \\ 
\footnotesize{$^{11}$\textit{ e-mail: \texttt{luca.schiavone[at]unina.it}}}  
}

\begin{document}

\maketitle

\begin{abstract}
The formulation of covariant brackets on the space of solutions to a variational problem is analyzed in the framework of contact geometry.
It is argued that the Poisson algebra on the space of functionals on fields should be read as a Poisson subalgebra within an algebra of functions equipped with a Jacobi bracket on a suitable contact manifold.
\end{abstract}

\section{Introduction}

The problem of describing a relativistic dynamical systems in a way which is manifestly covariant under the action of the Poincar\'{e} group has been repeatedly addressed, and many answers have been proposed.
Since the work by Peierls \cite{Peierls1952-Commutation_relativistic}, several proposals have been advanced in order to define a bracket for field theories which would overcome the equal-time formalism which has been traditionally followed.

The Hamiltonian description, which has been extremely useful in order to define quantum dynamical systems, seems to break the covariance from the very beginning  due to the choice of a splitting of the spacetime structure into space and time. 
However, a generalized Hamiltonian description has been developed starting from the 80's, where the role of symplectic geometry is replaced by the so called multisymplectic formalism (see, for instance, \cite{Gotay,Tulczyjew,CarinCrampIbort1991-Multisymplectic}, just to cite a few).
In this context, the works by Zuckerman,\cite{Zuckerman}, and Witten and Crnkovic,\cite{Witten}, plays a relevant role because, starting from the multisymplectic description, they introduced a Poisson bracket on the space of solutions of the associated de Donder-Weyl equations. 
This Poisson bracket is the departure point for a   quantum description of the associated quantum fields.

In the Lagrangian description, a different construction was already proposed some decades before.
In his pioneering work, Peierls  introduced a manifestly covariant bracket on the space of fields which are solutions of Euler-Lagrange equations. 
In his presentation, the action functional played a crucial role, but many geometrical aspects were not clearly analysed. 
Later on, this description was further developed by de Witt \cite{DeWitt1965-Groups_Fields}, who extended Peierls' construction in order to deal with gauge theories and more general fields theories. 

In this letter we want to outline a novel point of view on the geometrical structures which underlie the definition of  covariant brackets on the space of functionals on solutions of field equations. 
The main idea we want to convey in this letter is that the covariant bracket should be read in terms of a Poisson subalgebra  of the Jacobi algebra of functions on a  suitable contact manifold associated with the variational problem.
In particular, we will analyze in detail the case of non-relativistic Hamiltonian mechanics, and the relativistic particle.
In a companion letter \cite{C-DC-I-M-S-2020-02}, we will apply the ideas introduced here to the case of the Klein-Gordon fields, and the Schr\"{o}dinger equation seen as a field equation.
Since the aim of this letter consists in showing the connections between apparently different approaches, we will not focus too much on technical details, and reserve a more thorough exposition to future works.

\section{Non-relativistic Hamiltonian Mechanics}\label{sec: particle mechanics}

In this section, we consider the non-relativistic Hamiltonian mechanics of particles and use a geometric language which foreshadows future application to field theory.
Let $\mathcal{Q}$ be a smooth manifold, with local coordinates $(u^a)$, representing the configuration space of the particle.
In the non-relativistic framework, the existence of an {\itshape absolute simultaneity} leads us to consider the global ``time manifold'' $\mathcal{M} = \mathbb{R}$, and define the  extended configuration space $E = \mathcal{Q}\times\mathbb{R}$, with local coordinates $(u^a,s)$ and bundle projection $\pi_0\colon E\rightarrow\mathcal{M}$ locally given by $\pi_0(u^{a},s)=s$.

The manifold $\mathcal{P} = \mathbf{T}^*\mathcal{Q} \times \mathbb{R}$, where $\mathbf{T}^{*}\mathcal{Q}$ is the cotangent bundle of $\mathcal{Q}$ with local bundle coordinates $(u^{a},p_{a})$, is the extended phase space of the theory.
The extended phase space $\mathcal{P}$, with the associated coordinates\footnote{If necessary, we will assume these coordinates to be globally defined.}  $(u^a, p_a, s)$, is a trivial bundle with projection $\pi\,:\, \mathcal{P} \,\rightarrow \, \mathcal{M}$ given by 
\begin{equation}
\label{extended phase space}
\rho(u^a, p_a, s) = s
\end{equation} 
Given an Hamiltonian function $H \,:\, \mathcal{P} \ra \mathbb{R}$, we consider   the 1-form $\theta_{H}$ given by 
\begin{equation}
\theta_H = p_a \dd u^a - H \dd s\,,
\label{1-form theta_H}
\end{equation} 
where $p_{a}\dd u^{a}$ is the (pullback to the extended phase space of the) tautological form on $\mathbf{T}^{*}\mathcal{Q}$, and $\dd s $ is (the pullback to the extended phase space of) a volume form on the base manifold\footnote{For more details we refer to \cite{CarinCrampIbort1991-Multisymplectic}.}.
The manifold $\mathcal{P}$ is a contact manifold with contact 2-form   $\omega_H = \dd \theta_{H}$  (see \cite{AbraMars-Foundations_of_Mechanics,arnold-mathematical,C-C-M-2018} for the relevant definitions), and the  1-form $\theta_H$   is called contact 1-form. 
The 2-form $\omega_H$ possesses a family of vector fields in its kernel, called Reeb fields, which generates the so called characteristic leaves of the contact manifold. 
Since $\omega_H = \dd \theta_{H}$, it is possible to choose a particular Reeb vector field by fixing its value on $\theta_H$, i.e., by fixing a parametrization, and this will be relevant below when dealing with the equations of motions determining the dynamical trajectories of the system.

From a mathematical point of view, the fields of the theory are   sections $\chi$ of $\mathcal{P}$, that is, smooth maps such that $\pi\circ\chi=\mathrm{id}_{\mathcal{M}}$.
In coordinates, any such $\chi$ satisfies 
\be
\chi(s) = \left( u^a(s), p_a(s), s \right),
\ee
which clearly shows that the fields always come with a prescribed parametrization in terms of the parameter $s$.

The following diagram will be helpful to keep track of all the definitions introduced up to now.
\begin{center}
\begin{tikzcd}[row sep= huge, column sep= small]
E = \mathcal{Q}\times \mathbb{R} \arrow[dr, "\pi_0"] &  & \mathcal{P} = \mathbf{T}^*\mathcal{Q}\times \mathbb{R} \arrow[ll, "\pi_E"] \arrow[dl, "\pi" left]\\
& \mathcal{M} = \mathbb{R} \arrow[ur, dashed, bend right, "\chi" below] 
\end{tikzcd}
\end{center}

We write  $\mathscr{F}_{\mathcal{P}}$ for the space of all these sections, which  is not a smooth manifold in general, and a case-by-case analysis is required in order to equip it with the structure of a Banach or Hilbert manifold. 
Nevertheless,  in the following we will consider $\mathscr{F}_{\mathcal{P}}$ as a differential manifold from a  \grit{formal} point of view and we will use accordingly the notion of tangent vectors and differential forms. 
This will help in clarify the geometrical aspects of the constructions involved in the definition of the Action principle.

A {\itshape variation} for $\chi\in\Gamma$ is thought of as a tangent vector at $\chi$ as follows.
Given $\chi \in\mathscr{F}_{\mathcal{P}}$, consider a one-parameter family $\tilde{\chi} (v,s)$ of elements in $\mathscr{F}_{\mathcal{P}}$, with $v\in \left] -\epsilon, \epsilon \right[$, such that $\tilde{ \chi} (v=0,s) = \chi(s)$.
Two such families $\tilde{\chi}_{1}$ and $\tilde{\chi}_{2}$ are defined to be equivalent if their ``derivatives'' with respect to $v$ coincides at $v=0$.
Then, a variation of $\chi$ is defined as an equivalence class of ``tangent vectors'' at $v=0$ that can be written as 
\begin{equation}
U_{\chi}= (\delta u^a(s), \delta p_a(s), s)\,=\, \frac{\partial}{\partial v}\left(\tilde{\chi}\right)_{v=0} \,,
\end{equation}
and thus identifies a vector field 
\be
U_{\chi} =  \delta u^a(s) \frac{\partial }{\partial u^a} + \delta p_a(s)\frac{\partial }{\partial p_a}
\ee
along the section $\chi(s)$ which is vertical with respect to the fibration $\pi\colon\mathcal{P}\ra\mathcal{M}$.
The tangent space $\mathbf{T}_{\chi}\mathscr{F}_{\mathcal{P}}$ at $\chi$ is given by all the $U_{\chi}$.
In the following, it will be useful to extend $U_{\chi}$  to a vertical vector field $\tilde{U}$ in a neighbourhood of the image of $\chi$ inside $\mathcal{P}$.


Now, we pass to describe the dynamics in terms of the (Schwinger-Weiss) action principle.
Given a Hamiltonian function $H$, we build $\theta_H$ as in Eq. \eqref{1-form theta_H}, and the action functional   $S\colon\mathscr{F}_{\mathcal{P}}\lra\mathbb{R}$ given by
\begin{equation}
S[\chi] = \int_{\mathcal{M}} \chi^*(\theta_{H}) = \int_{\mathcal{M}}\left( p_a\frac{\dd u^a}{\dd s} - H \right) \mathrm{d}s =: \int_{\mathcal{M}} \lag(H) \dd s \,.
\end{equation}   
Given  $U_{\chi}\in\mathbf{T}_{\chi}\mathscr{F}_{\mathcal{P}}$, the variation\footnote{In case a Banach space structure for $\mathscr{F}_{\mathcal{P}}$ exists, $\mathrm{d}S$ can be thought of as a proper differential, otherwise, it should be read as  formal differential.} $\mathrm{d}S[\chi](U_{\chi})$ of $S$ with respect to $\delta\chi$ at $\chi$ is nothing but  \cite{Saunders1989-Jet_bundles}
\be
\begin{split}
\mathrm{d}S[\chi](U_{\chi})&=  \frac{\partial}{\partial v} S[\tilde{\chi} ] |_{v=0} = \int_{\mathcal{M}}\chi^*(\mathrm{L}_{\tilde{U}}\theta_{H}) = \int_{\mathcal{M}}\chi^*(i_{\tilde{U}})\dd \theta_{H} + \int_{\mathcal{M}}\chi^*(\dd i_{\tilde{U}}\theta_{H}),
\end{split}
\ee
where $\tilde{U}$ is any extension of $U_{\chi}$.
Exploiting Stokes' theorem, we have
\be
\mathrm{d}S[\chi](U_{\chi})=\int_{\mathcal{M}}\chi^*(i_{\tilde{U}}\dd \theta_{H}) +\int_{\partial\mathcal{M}}i_{\partial\mathcal{M}}^{*}\chi^*( i_{\tilde{U}}\theta_{H})\,=\,\int_{\mathcal{M}}\chi^*(i_{\tilde{U}}\dd \theta_{H})
\ee
because the boundary term vanishes being $\partial\mathbb{R}=\emptyset$.
The action principle  states that the dynamical trajectories satisfy the Euler-Lagrange equations
\begin{equation}\label{eqn: Schwinger-Weiss action principle 1}
\mathbb{EL}_{\chi}(U_{\chi}) := \int_{\mathcal{M}}\chi^*(i_{\tilde{U}}\dd \theta_{H}) = 0 \,,\quad \forall U_{\chi} \in \mathbf{T}_{\chi}\mathscr{F}_{\mathcal{P}}\,,
\end{equation}
that can be geometrically interpreted as the contraction of the 1-form $\mathbb{EL}$ with a tangent vector $U_{\chi}\in \mathbf{T}_{\chi}\mathscr{F}_{\mathcal{P}}$.
In the case of a field theory in which the base space $\mathcal{M}$ has a non-trivial boundary, we have
\be\label{eqn: Schwinger-Weiss action principle}
\mathrm{d}S_{\chi}(U_{\chi})=\mathbb{EL}_{\chi}(U_{\chi}) +\int_{\partial\mathcal{M}}i_{\partial\mathcal{M}}^{*}\chi^*( i_{\tilde{U}}\theta_{H}),
\ee
and the equations of motions are obtained by considering the action principle in the  Schwinger-Weiss form, where the variation of the action functional at a solution is assumed to depend only on the value of the fields at the boundary.
We also note that such a geometrical reformulation of the Schwinger-Weiss variational principle could be useful to introduce a Quantum Action Principle in the groupoid reformulation of Schwinger's algebra of selective measurements which has been recently proposed by some of the authors (see \cite{C-I-M-2018,C-I-M-02-2019,C-I-M-03-2019,C-I-M-05-2019,C-DC-I-M-2020,C-DC-I-M-02-2020} for more details).

The space of dynamical trajectories is denoted by $\mathcal{EL}_{\mathcal{M}}$.
From Eq. \eqref{eqn: Schwinger-Weiss action principle 1}, the dynamical trajectories are easily seen to satisfy Hamilton equations
\be\label{eqn: Ham eq}
\begin{split}
\frac{\dd s }{\dd s}\,=\,1 ,\quad \frac{\dd u^{a} }{\dd s} \,=\,\frac{\partial H}{\partial p_a},\quad
\frac{\dd p_{a} }{\dd s}\,=\,-\frac{\partial H}{\partial u^a}, 
\end{split}
\ee
where the first equation follows from the requirement for $\chi$ to be a section of $\pi\colon\mathcal{P}\ra\mathcal{M}$.
As we will now see, $\mathcal{EL}_{\mathcal{M}}$ is endowed with a (pre)symplectic structure.
Let us fix $s_0\in \mathcal{M}$ and think of $\Sigma\equiv\{s_{0}\}$ as a codimension-one submanifold of $\mathcal{M}$ embedded  via the canonical immersion $i_{\Sigma}$. 
Then, we define the  2-form $\Omega^{s_{0}}$ on $\mathcal{EL}_{\mathcal{M}}$ given by
\begin{equation}
\Omega_{\chi}^{s_0}(U_{\chi},V_{\chi}) = \int_{\Sigma} i_{\Sigma}^*\chi^*(i_{\tilde{V}}i_{\tilde{U}} \dd \theta_{H}) =  \delta u^a_{U}(s_{0})\,\delta p_a^{V}(s_{0}) - \delta u^a_{V}(s_{0})\,\delta p_a^{U}(s_{0})\,,
\end{equation}
where $\chi\in\mathcal{EL}_{\mathcal{M}}$ and\footnote{The formal tangent space $\mathbf{T}_{\chi}\mathcal{EL}_{\mathcal{M}}$ may be defined in analogy with $\mathbf{T}_{\chi}\mathscr{F}_{\mathcal{P}}$ by considering one-parameter families of solutions in $\mathcal{EL}_{\mathcal{M}}$.} $U_{\chi},V_{\chi} \in \mathbf{T}_{\chi}\mathcal{EL}_{\mathcal{M}}$. 
We want to show that, since $\chi$ belongs to the submanifold $\mathcal{EL}_{\mathcal{M}}\subset \mathscr{F}_{\mathcal{P}}$, $\Omega^{s_0}_{\chi}$ is actually independent of the choice of a specific $\Sigma \subset \mathcal{M}$. 
Indeed, we may consider $s_{0}<s_{1}\in\mathcal{M}$, and take $\mathcal{M}_{01}=[s_{0},s_{1}]$ as the base manifold of our theory  so that we have
\begin{equation}
\Omega_{\chi}^{s_1}(U_{\chi},V_{\chi}) - \Omega_{\chi}^{s_0}(U_{\chi},V_{\chi}) = \int_{\partial \mathcal{M}_{01}} i_{\partial \mathcal{M}_{01}}^*\chi^*(i_{\tilde{V}}i_{\tilde{U}} \dd \theta_{H}) .
\end{equation} 
Then, Eq. \eqref{eqn: Schwinger-Weiss action principle} implies
\be
\mathrm{d}S_{\chi}(U_{\chi})=\mathbb{EL}_{\chi}(U_{\chi})  + \int_{\partial\mathcal{M}_{01}}\,i_{\partial\mathcal{M}_{01}}^{*}\chi^*( i_{\tilde{U}}\theta_{H}),
\ee
and thus
\be
-\mathrm{d}\mathbb{EL}_{\chi}(U_{\chi},V_{\chi})=  \int_{\partial\mathcal{M}_{01}}\,i_{\partial\mathcal{M}_{01}}^{*}\chi^*( i_{\tilde{U}} i_{\tilde{V}}\mathrm{d}\theta_{H})=\,\Omega_{\chi}^{s_1}(U_{\chi},V_{\chi}) - \Omega_{\chi}^{s_0}(U_{\chi},V_{\chi}).
\ee
However, if $\chi$ is a dynamical trajectory then  $\mathrm{d}\mathbb{EL}_{\chi}(U_{\chi},V_{\chi})=0$ and thus
\begin{equation}
\Omega_{\chi}^{s_1}(U_{\chi},V_{\chi}) - \Omega_{\chi}^{s_0}(U_{\chi},V_{\chi}) =   0
\end{equation} 
as claimed.
Consequently, we will write $\Omega$ instead of $\Omega^{s_{0}}$.

If $\Omega$ is invertible, then we obtain a Poisson bracket on the space of functionals on $\mathcal{EL}_{\mathcal{M}}$,  which is covariant in the sense that it transforms covariantly under the action of the symmetry transformations   preserving $\mathcal{EL}_{\mathcal{M}}$.
However, the bracket associated with $\Omega$ is difficult to compute precisely because it requires us to invert $\Omega$.
For instance, Peierl's approach \cite{Peierls1952-Commutation_relativistic} always requires to find the advanced and retarded Green functions of some system of differential equations \cite{AsoCiagliaDCosmoIbort2017-Covariant_brackets,B-F-R-2019,DH-F-R-W-2020,Fredenhagen2015-Algebraic_quantum_field_theory}.
Here, we want to provide an alternative way to realize the bracket associated with $\Omega$ which relies on the contact structure of the extended phase space.
Specifically, we will see that $\mathcal{EL}_{\mathcal{M}}$ may be identified with a suitable quotient space of the extended phase space, in such a way that the functionals on $\mathcal{EL}_{\mathcal{M}}$ are identified with a particular subalgebra of smooth functions on the extended phase space.
This subalgebra will then be proved to inherit  a  Poisson bracket coming from the  Jacobi bracket naturally associated with the contact structure \cite{AsoCiagliaDCosmoIbortMarmo2017-Covariant_Jacobi_brackets,C-C-M-2018}.
This Poisson bracket is precisely the Poisson bracket associated with $\Omega$.

At this purpose, we start noting that the dynamical trajectories satisfying Eq. \eqref{eqn: Ham eq} can  be described as the integral curves of the vector field $X_{H}$  given by
\begin{equation}
X_{H} = \frac{\partial }{\partial s} + \frac{\partial H}{\partial p_a} \frac{\partial }{\partial u^a} - \frac{\partial H}{\partial u^a}\frac{\partial }{\partial p_a}
\end{equation}
on $\mathcal{P}$ satisfying 
\begin{equation}
i_{X_H}\dd \theta_{H} = 0\,.
\end{equation}
This means that the vector field $X_H$  is in the kernel of the contact 2-form $\dd \theta_H$ and determines a particular parameterization, in terms of the evolution parameter $s$, of the characteristic foliation of the contact manifold.
It must be noted, indeed, that any vector field of the form $f\,X_{H}$, with $f$ a smooth, non-vanishing function on $\mathcal{P}$ is again in the kernel of $\dd \theta_{H}$, and the support of its integral curves coincide with the support of the integral curves of $X_{H}$.
Accordingly, we may interpret the integral curves of $f\,X_{H}$ as reparametrizations of the dynamical trajectories.
The particular choice\footnote{It may happen that this condition may be fulfilled only on an open submanifold of $\mathcal{P}$, for instance, if there are degenerate dynamical trajectories  consisting of points.
Furthermore, if $\mathcal{P}=\mathbf{T}^{*}\mathcal{Q}\times\mathbb{R}$ is replaced by a manifold in which the contact two-form is not exact, like it happens for the unitary evolutions of pure quantum states\cite{C-DC-L-M-2017,C-DC-I-L-M-2017} where $\mathbf{T}^{*}\mathcal{Q}$ would be replaced by the complex projective space $\mathbb{CP}(\hh)$, the choice of the normalization of $X_{H}$ must be dealt with differently.} $i_{\Gamma}\theta_H = 1$ defines what is usually called Reeb vector field. Let us notice that $i_{X_H}\theta_H = \lag(H)$, therefore the integral curves of the Reeb vector field $\Gamma$ will not be sections of $\pi\colon\mathcal{P}\ra\mathcal{M}$ anymore.  
Under suitable regularity properties for $X_H$, the family of its integral curves   defines a regular foliation of the manifold $\mathcal{P}$, and the associated quotient space, say $\mathcal{N}_H$, inherits a differential structure. 
Furthermore,  it should be clear now that every point in $\mathcal{N}_H$ can be identified with one and only one element in $\mathcal{EL}_{\mathcal{M}}$ as anticipated before, and thus we see that the space of dynamical trajectories inherits the structure of a finite-dimensional smooth manifold.

Now, we may look at the algebra of smooth functions on $\mathcal{N}_H$ as a subalgebra of the algebra of functions on $\mathcal{P}$, specifically, the algebra  $C^{\infty}_{H}(\mathcal{P})$ of functions $f\in C^{\infty}(\mathcal{P})$ such that $\mathrm{L}_{X_H}f=0$. Since we are assuming that $\mathcal{N}_H$ is diffeomorphic to $\mathcal{EL}_{\mathcal{M}}$, the algebra $C^{\infty}(\mathcal{N}_H)$ will be isomorphic to the algebra $C^{\infty}(\mathcal{EL}_{\mathcal{M}})$.
Then, on this realization of the algebra of smooth functions on $\mathcal{N}_{H}$ is possible to describe the Poisson bracket associated with $\Omega$ in terms of the Jacobi bracket on $\mathcal{P}$  naturally associated  with the contact structure.
Indeed, since $\mathcal{P}$ is a contact manifold, there is Lie bracket $\left[ \cdot , \cdot \right]_J$ on the space of smooth functions,  called  Jacobi bracket, which   is the contact counterpart of a Poisson bracket on a symplectic manifold  \cite{AsoCiagliaDCosmoIbortMarmo2017-Covariant_Jacobi_brackets,C-C-M-2018}.
The Jacobi bracket may be written in terms of a bivector $\Lambda$ and the Reeb vector field $\Gamma$ as  
\begin{equation}
\left[ f , g \right]_J = \Lambda(\dd f , \dd g) + f\mathrm{L}_{\Gamma}g - g\mathrm{L}_{\Gamma}f\,.
\label{jacobi_bracket}
\end{equation}   
Referring to the so-called generalized Darboux coordinates \cite{AbraMars-Foundations_of_Mechanics} $(Q^{a},P_{a},W)$   according to which the contact 1-form can be written as $\theta_H = \dd W + P_a \dd Q^a$, the Reeb vector field becomes $\Gamma = \frac{\partial}{\partial W}$, and the bivector $\Lambda$ is expressed as
\begin{equation}
\Lambda = \left( \frac{\partial }{\partial Q^a} - P_a \frac{\partial}{\partial W} \right) \wedge \frac{\partial}{\partial P_a}   \,.
\end{equation}
Let us notice that it is possible to add any function of $(Q^a,P_a)$ to $W$ without altering the form, and similarly we may perform any canonical transformation on the variables $(Q^a, P_a)$. Even if the Jacobi bracket  satisfies the Jacobi identity, it does not give rise to derivations for the product of functions, differently from what happens for a Poisson bracket.
Given a function $f \in C^{\infty}(\mathcal{P})$, there is an associated ``Jacobian'' vector field $X_f = \Lambda(\dd f,\,\cdot) + f \Gamma$ such that
\begin{equation}
\left[ X_f\,,X_g \right] = X_{\left[ f,g\right]_J}\,.
\end{equation} 
Moreover, the (pointwise)  subalgebra $C^{\infty}_H(\mathcal{P})$  is also a Lie subalgebra with respect to the Jacobi bracket given in Eq. \eqref{jacobi_bracket}, and it is straightforward to prove that  
\begin{equation}
\left[ f,g \right]_J = \Lambda(\dd f, \dd g)\,,
\end{equation} 
for $f,g \in C^{\infty}_H(\mathcal{P})$, so that it is actually a Poisson subalgebra in the sense that the bracket defines derivations of the pointwise product in  $C^{\infty}_H(\mathcal{P})$.
This bracket on $C^{\infty}_H(\mathcal{P})$ represents the Poisson bracket associated with $\Omega$ on $\mathcal{EL}_{\mathcal{M}}$ from the unfolded point of view.

On the other hand, there is another way to describe the Poisson bracket on functions on $\mathcal{EL}_{\mathcal{M}}$ which is linked to an identification of $\mathcal{N}_{H}$ inside  $\mathcal{P}$.
Specifically, let us assume for simplicity that the generalized Darboux coordinates $(Q^{a},P_{a},W)$ are globally defined so that the level set $\mathcal{W} :=W^{-1}(c_{0})$, with $c_{0}\in\mathbb{R}$, is a submanifold of $ \mathcal{P}$ which is diffeomorphic to\footnote{If the Darboux coordinates are not global, one should look for a foliation which is transversal to the characteristic foliation associated with $\omega$, and such that its leaves are symplectic submanifolds.} $\mathbf{T}^*\mathcal{Q}$.
Furthermore, $\mathcal{W}$ can be identified with the space $\mathcal{N}_{H}$ because, in this new set of coordinates, $X_{H}$ is transversal to $\mathcal{W}$  and  the dynamical trajectories $\chi$ may be labelled by  points in $\mathcal{W}$. 
Therefore, the space of solutions $\mathcal{EL}_{\mathcal{M}}$ can be identified with $\mathcal{W}$  which is a symplectic submanifold of the contact manifold $\mathcal{P}$, the symplectic form being the pullback of the contact 2-form to $\mathcal{W}$. 
Now, given a function $f \in C^{\infty} (\mathcal{W})\cong C^{\infty}(\mathcal{N}_{H})\cong C^{\infty}_H(\mathcal{P})\cong C^{\infty}(\mathcal{EL}_{\mathcal{M}}) $ we can define the  vector field  
\begin{equation}
\mathbf{X}_f = \Lambda_{\mathcal{W}}(\dd f\, ,\cdot )\,,
\label{peierls hamiltonian vector fields}
\end{equation} 
where $\Lambda_{\mathcal{W}}$ is the inverse of the symplectic form on $\mathcal{W}$.
Then, a Poisson bracket on $C^{\infty} (\mathcal{W})\cong C^{\infty}(\mathcal{EL}_{\mathcal{M}}) $  is easily given by
\be
 \left\lbrace f, g \right\rbrace_W   = \Lambda_{\mathcal{P}}(\dd f\, ,\dd g )=\mathrm{L}_{\mathbf{X}_f} g.
\ee
It is not hard to see that 
\be
i_{\mathcal{W}}^{*}\left([f,\,g]_{J}\right)\,=\,  \left\lbrace i_{\mathcal{W}}^{*}f, i_{\mathcal{W}}^{*}g \right\rbrace_W,
\ee
where $f,g\in C^{\infty}_H(\mathcal{P})$, and $i_{\mathcal{W}}\colon\mathcal{W}\lra\mathcal{P}$ is the canonical immersion, meaning that the two brackets agree.
However, note that the bracket $ \left\lbrace \cdot, \cdot \right\rbrace_W $  depends on the particular choice of the function $W$ and, therefore, on the identification of $\mathcal{W}$ with $\mathcal{N}_{H}$, and thus introduces an arbitrariness, while the bracket $[\cdot,\cdot]_{J}$ emerges  naturally from the contact structure on the extended phase space $\mathcal{P}$. This identification leads to a decomposition of the space $\mathfrak{X}(\mathcal{P})$ of vector fields on $\mathcal{P}$ into the direct sum 
\begin{equation}
\mathfrak{X}(\mathcal{P}) = \mathfrak{X}^v(\mathcal{P})\oplus \mathfrak{X}^h(\mathcal{P})
\end{equation}
of vertical and horizontal vector fields, which is the choice of a connection of the bundle $\pi\colon\mathcal{P}\rightarrow\mathcal{M}$. As we have said, this choice is not univoquely determined. 
Indeed, if we choose a different system of coordinates, namely:
\begin{equation}
\tilde{Q}^a = Q^a\,,\quad \tilde{P}_a = P_a\,,\quad \tilde{W}= W + \frac{P_aQ^a}{2}
\end{equation}
the contact 1-form $\theta_H$ and 2-form $\omega_H$ as well as the Reeb field $\Gamma$ and the bivector field $\Lambda$ are expressed as follows:
\begin{eqnarray*}
& \theta_H = \dd \tilde{W} - \frac{\tilde{P}_a}{2}\dd \tilde{Q}^a + \frac{\tilde{Q}^a}{2}\dd \tilde{P}_a\,,\quad \omega_H = \dd \tilde{P}_a\wedge \dd \tilde{Q}^a  \\
& X_H = \frac{\partial}{\partial \tilde{W}} \,,\quad \Lambda = \left( \frac{\partial}{\partial \tilde{Q}}^a - \frac{\tilde{P}_a}{2}\frac{\partial}{\partial \tilde{W}} \right)\wedge \left( \frac{\partial}{\partial \tilde{P}_a} + \frac{\tilde{Q}^a}{2}\frac{\partial}{\partial \tilde{W}} \right)\,. 
\end{eqnarray*}
Therefore, the level set of the function $\tilde{W}$ are transversal to the vector field $X_H$ and one can identify the quotient space $\mathcal{N}_{H}$ with the submanifold $\mathcal{P}\supset\tilde{\mathcal{W}}=\tilde{W}^{-1}(c_0)$, for some $c_0\in \mathbb{R}$. The pullback of the 2-form to $\tilde{\mathcal{W}}$ is the symplectic form $\tilde{\omega}_H = \dd \tilde{P}_a \wedge \dd \tilde{Q}^a$ and the bivector field associated with the Poisson bracket on the algebra of functions $C^{\infty}(\tilde{\mathcal{W}})$ is 
\begin{equation}
\tilde{\Lambda}_{\mathcal{W}} = \frac{\partial}{\partial \tilde{P}_a}\wedge\frac{\partial}{\partial \tilde{Q}_a}\,.
\end{equation}
It is immediate to notice that this bivector field is not tangent to the level set of the function $W = \tilde{W} - \frac{\tilde{P}_a\tilde{Q}^a}{2}$, showing explicitly the dependence of the lift 
\begin{equation}
\mathbf{X}_f = \tilde{\Lambda}_{\mathcal{W}}(\dd f\,,\cdot)
\end{equation}
on the identification of $\mathcal{N}_{H}$ with the different submanifold $\tilde{W}\subset \mathcal{P}$. On the other hand it is still true that 
\begin{equation}
i_{\tilde{\mathcal{W}}}^{*}\left([f,\,g]_{J}\right)\,=\,  \left\lbrace i_{\tilde{\mathcal{W}}}^{*}f, i_{\tilde{\mathcal{W}}}^{*}g \right\rbrace_{\tilde{W}},
\end{equation}
where $f,g\in C^{\infty}_H(\mathcal{P})$, and $i_{\tilde{\mathcal{W}}}\colon\tilde{\mathcal{W}}\lra\mathcal{P}$ is the canonical immersion.

The following diagram will pictorially summarize the previous discussion.
\begin{center}
\begin{tikzcd}[row sep = huge, column sep = small]
\mathbf{T}^*\mathcal{Q}\simeq \mathcal{W} \arrow[rr,hook, "i_{\mathcal{W}}"] \arrow [dr, leftrightarrow, dashed] & & \mathcal{P}=\mathbf{T}^*\mathcal{Q}\times \mathbb{R} \arrow [dl, "X_H"] \arrow[dr, "{\frac{\partial}{\partial s}}" below ]& & i_{\Sigma}^*(\mathcal{P})\simeq \mathbf{T}^*\mathcal{Q} \arrow[ll, hook' , "i_{\Sigma}^*(\pi)" above] \arrow [dl, leftrightarrow, dashed]\\
\mathcal{EL}_{\mathcal{M}} \arrow[r, phantom, "\simeq" description] &\mathcal{N}_H & & \mathbf{T}^*\mathcal{Q} 
\end{tikzcd}
\end{center}

The extended phase space $\mathcal{P}$ can be foliated with respect to the action of the vector fields $X_H$ associated with the dynamics, and $\frac{\partial}{\partial s}$ associated with the absolute simultaneity on $E$. The two quotients are $\mathcal{N}_H$ and $\mathbf{T}^*\mathcal{Q}$, respectively. The latter can be identified with the space of fields on a codimension-one submanifold of the base manifold $\mathcal{M}$. The former, which represents the space of solutions, can be identified with a submanifold $\mathcal{W}\subset \mathcal{P}$ of constants of the motion. This submanifold corresponds to a ``dynamical simultaneity surface'' for a chosen ``dynamical time'' $W$. This choice allows us to define a Poisson bracket on the space of function $C^{\infty}(\mathcal{W})\cong C^{\infty}(\mathcal{N}_H)$, using the symplectic structure which is the pullback of $\omega_H$ to $\mathcal{W}$. However, $C^{\infty}(\mathcal{N}_H)$ is isomorphic to $C^{\infty}_H(\mathcal{P})$, which is a subalgebra of $C^{\infty}(\mathcal{P})$ with respect to the Jacobi bracket defined on the exact contact manifold $\mathcal{P}$. This subalgebra is actually a Poisson subalgebra, and in this case the definition of the bracket does not depend on the choice of a section of the quotient manifold $\mathcal{N}_H$ within $\mathcal{P}$.  

\begin{example}
As an example of the whole construction let us consider the motion of a free particle on a line. In this case the configuration manifold is $\mathcal{Q}=\mathbb{R}$ and the Hamiltonian function is $H = \frac{1}{2}p^2$. If we exclude the zero section $p=0$ in $\mathcal{P}= \mathbf{T}^*\mathbb{R}\times \mathbb{R}$, the contact structure is defined via the differential forms:
\begin{equation}
\theta_H = p \dd q - \frac{1}{2}p^2 \dd s  \,,\quad \omega_H = \dd \theta_H = \dd p \wedge \dd q - p \dd p \wedge \dd s\,,
\end{equation} 
and the Hamiltonian vector field $X_H$ determining the dynamics is:
\begin{equation}
X_H = \left( \frac{\partial }{\partial s} + p \frac{\partial }{\partial q}\right) \,.
\end{equation}
The Darboux coordinates are:
\begin{equation}
W = \frac{p^2}{2} s\,,\quad Q=q-ps\,,\quad P=p\,,
\end{equation} 
which are obtained via the canonical transformation generated by the solution $S(Q,q,s)$ of the Hamilton-Jacobi equation $\frac{\partial S}{\partial s} + H(q, \frac{\partial S}{\partial q})=0$ 
\begin{equation}
S = \frac{(q-Q)^2}{2s}\,.
\end{equation}

The Jacobi bracket can be written in terms of the bivector field 
\begin{equation}
\Lambda = \left( \frac{\partial }{\partial Q} - P\frac{\partial }{\partial W} \right) \wedge \frac{\partial }{\partial P}
\end{equation}
and the Reeb field $\Gamma = \frac{2}{p^2}X_H$, which in the new coordinates reduces to $\Gamma=\frac{\partial}{\partial W}$. The space of solutions can be identified, for instance, with the submanifold $\mathcal{W}$ defined by the condition $W=0$, which is a symplectic manifold with the symplectic structure given by $\tilde{\omega} = \dd P \wedge \dd Q$. 

In this new set of coordinates the flow associated to the vector field is the identity map and the vector field $\mathbf{X}_f$ associated with any function $f\in C^{\infty}(\mathcal{N}_{H})$, is constant along a solution of the Hamilton equations. Then
\begin{equation}
\mathbf{X}_f = \frac{\partial f}{\partial P}\frac{\partial }{\partial Q} - \frac{\partial f}{\partial Q}\frac{\partial}{\partial P}
\end{equation}
the  Poisson bracket $\left\lbrace  f, g\right\rbrace_W $ is simply $\mathrm{L}_{\mathbf{X}_f} g$.
\end{example}

\section{Relativistic particle}

A different situation emerges if one considers the dynamics of a relativistic particle.
Indeed, in the relativistic context, the extended phase space the way we have considered in previous section is not available because of the absence of an absolute notion of simultaneity, i.e., it is not possible to select a ``time function'' projecting on $\mathbb{R}$ while preserving Poincar\'{e} invariance.
For a massive particle, the relativistic counterpart of the extended phase space is given by the mass-shell submanifold.
Then, the choice of a fibration  of the mass-shell submanifold reflects the choice of a reference frame which, as we will see, may be ``kinematical'' or ``dynamical''.

Let us consider Minkowski spacetime $E=(\mathbb{R}^4,\eta)$, where $\eta = \dd u^0\otimes \dd u^0- \delta_{jk}\dd u^j\otimes\dd u^k$ is the Minkowski metric tensor with $(u^{0},u^{1},u^{2},u^{3})$ a global Cartesian coordinates system. 
The cotangent bundle $\mathbf{T}^{*}E$, with global Cartesian coordinates $(u^{\mu}, p_{\mu})$, is interpreted as the energy-momentum bundle  for a single particle, and, as said before, to deal with a specific particle, say a spinless particle with mass $m$, we must consider positions and momenta satisfying the mass-shell relation\footnote{In the following, we will focus only on the connected component of $\Sigma_{m}$ individuated by $p_{0}>0$, and, with  an evident abuse of notation, we will keep denoting it by $\Sigma_{m}$.} 
\begin{equation}\label{eqn: mass shell}
\phi(u^{\mu}, p_{\mu}) = \eta^{\mu \nu}p_{\mu} p_{\nu} = m^2\,,
\end{equation}
defining a submanifold  $\mathcal{P}$ of $\mathbf{T}^*E$.
Note that this submanifold is invariant with respect to the cotangent lift of the canonical action of the Poincar\'{e} group on $E$.

The relativistic analogue of the extended phase space $\mathbf{T}^{*}\mathcal{Q}\times\mathbb{R}$ is precisely the mass-shell   $\mathcal{P} \subset\mathbf{T}^{*}E$, and the fibration of $\mathcal{P}$ onto $\mathcal{M} = \mathbb{R}$ depends on the choice of a {\itshape time function} $\tau\colon\mathcal{P}\lra\mathcal{M}$.
For the moment, we will make use of a {\itshape kinematical time function}, which is defined in terms of a   time function  in the sense of general relativity, that is, a smooth function $t\colon E \ra\mathcal{M}$ giving rise to a spacetime splitting where the spacelike leaves are the level sets of $t$.
Note that a time function exists on every {\itshape causal spacetime}, and, in particular, on every globally hyperbolic spacetime like Minkowski spacetime.
Then, we define the  {\itshape kinematical time function} $\tau\colon\mathcal{P}\ra\mathcal{M}$ setting
\be
\tau:=\pi_{\mathcal{P}}\circ t,
\ee
where $\pi_{\mathcal{P}}$ is the pullback to $\mathcal{P}$ of the canonical cotangent bundle projection $\pi\colon\mathbf{T}^{*}E\ra E$.
For instance, if $(u^{0},u^{1},u^{2},u^{3})$ is the set of global Cartesian coordinates on $E$ given above, we may set $t:=u^{0}$, so that $\tau(u^{\mu},p_{\mu})\,=\,u^{0}$.
It is not hard to see that this choice of the time function $\tau$ gives a diffeomorphism  between $\mathcal{P}$ and $\mathbf{T}^{*}\mathbb{R}^{3}\times\mathbb{R}$ which, as we will see below,  leads to the description of the relativistic particle {\itshape \`{a} la} Landau.

The fields are sections of  $\tau\colon\mathcal{P}\ra\mathcal{M}$   and the space of all such fields is denoted by $\mathscr{F}_{\mathcal{P}}$.
For instance, every $\chi\in\mathscr{F}_{\mathcal{P}}$ is locally written as
\be
\chi(s)\,=\,(s,u^{j}(s),p_{\mu}(s))\,,
\ee
which clearly shows that the fields always come with a prescribed parametrization in terms of the parameter $s$.

Now, just as we did in the previous section, we identify a variation along $\chi$   with a vector field $U_{\chi}$ along $\chi$ itself, which is vertical with respect to $\tau\colon\mathcal{P}\lra \mathcal{M}$
\be
U_{\chi}\,=\,\delta u^{j}(s)\,\frac{\partial}{\partial u^{j}} + \delta p_{\mu}(s)\frac{\partial }{\partial p_{\mu}},
\ee
and, because of the mass shell constraint given in Eq. \eqref{eqn: mass shell}, must satisfy 
\be
\eta^{\mu\nu}\delta p_{\mu}(s)\,p_{\nu}(s)\,=\,0
\ee
for all admissible variations $\delta p_{\mu}(s)$. 

The pullback to $\mathcal{P}$ of the canonical symplectic form $\omega=\dd p_{\mu}\wedge\dd u^{\mu}$ on $\mathbf{T}^{*}E$ determines a contact 2-form denoted by $\omega_m$, while the pullback to $\mathcal{P}$ of the  tautological 1-form $\theta=p_{\mu}\dd u^{\mu}$ on $\mathbf{T}^{*}E$ determines a contact 1-form denoted by $\theta_{m}$.
Then, it should not come as a surprise that the role of $\theta_{H}$ in this relativistic context is taken by the 1-form $\theta_{m}$, so that the action functional reads
\be
S[\chi]\,:=\,\int_{\mathcal{M}}\,\chi^{*}\theta_{m}\,.
\ee
Proceeding as in section \ref{sec: particle mechanics}, we obtain
\be
\mathrm{d}S[\chi](U_{\chi})=\int_{\mathcal{M}}\chi^*(i_{\tilde{U}} \dd \theta_{m}) +\int_{\partial\mathcal{M}}i_{\partial\mathcal{M}}^{*}\chi^*( i_{\tilde{U}}\theta_{m})\,=\,\int_{\mathcal{M}}\chi^*(i_{\tilde{U}} \dd \theta_{m}),
\ee
because $\partial\mathcal{M}=\emptyset$.
Therefore, the Schwinger-Weiss action principle allows us to characterize the set of dynamical trajectories $\mathcal{EL}_{\mathcal{M}}$ as the set of all $\chi\in\mathscr{F}_{\mathcal{P}}$ satisfying
\be
\mathbb{EL}_{\chi}(U_{\chi})\,=\,\int_{\mathcal{M}}\chi^*(i_{\tilde{U}} \dd \theta_{m})\,=\,0\;\;\;\forall\,U_{\chi}\,\in\mathbf{T}_{\chi}\mathscr{F}_{\mathcal{P}}.
\ee
An explicit computation shows that the dynamical trajectories must satisfy
\be
\begin{split}
\frac{\dd u^{0} }{\dd s}\,=\,1 ,\quad \frac{\dd u^{j} }{\dd s} \,=\,-\frac{\delta^{jk}p_{k}}{p_{0}},\quad
\frac{\dd p_{\mu} }{\dd s}\,=\,0. 
\end{split},
\ee
As done in section \ref{sec: particle mechanics}, considering $\Sigma=\{s_{0}\}\subset\mathcal{M}$  as a codimension-one submanifold of the base manifold of the fibration $\tau\colon\mathcal{P}\ra\mathcal{M}$, it is possible to define  the  2-form $\Omega^{s_{0}}$ on $\mathcal{EL}_{\mathcal{M}}$ given by
\begin{equation}
\Omega_{\chi}^{s_0}(U_{\chi},V_{\chi}) = \int_{\Sigma} i_{\Sigma}^*\chi^*(i_{\tilde{V}}i_{\tilde{U}} \dd \theta_{H}) =  \delta u^{j}_{U}(s_{0})\,\delta p_{j}^{V}(s_{0}) - \delta u^{j}_{V}(s_{0})\,\delta p_{j}^{U}(s_{0})\,,
\end{equation}
and it is possible to show that $\Omega^{s_{0}}$ actually does not depend on the choice of $s_{0}$ following the argument given in section \ref{sec: particle mechanics}.

The dynamical trajectories can be identified with the integral curves of the vector field
\be
X\,=\,\frac{\partial}{\partial u^{0}} - \frac{\delta^{jk}p_{k}}{p_{0}} \frac{\partial}{\partial u^{j}}
\ee
on $\mathcal{P}$ satisfying
\begin{equation}
i_{X}\omega_m = 0\,.
\end{equation}
This is precisely the description of the relativistic particle done by Landau and Lifshitz \cite{landau-lifshitz}.
Indeed, by solving the mass-shell constraint in Eq. \eqref{eqn: mass shell} with respect to $p_{0}$, we obtain 
\be
p_{0}\,=\,\sqrt{\delta^{jk}\,p_{j}p_{k} + m^{2}},
\ee
and thus, setting $H:=p_{0}$ and $s=u^{0}$,  $\theta_{m}$ reads
\be
\theta_{m}\,=\,p_{j}\,\dd x^{j} + H\,\dd s ,
\ee
so that we can re-interpret the system as a Hamiltonian system with Hamiltonian given by $H=p_{0}$.
However, it is important to stress that this Hamiltonian interpretation is tied to the choice of the kinematical time function $\tau=u^{0}$.
As we will now see, a different choice of time function leads to a different description of the system.

Now, we will describe the Poisson bracket on the space of functionals on the space of solutions $\mathcal{EL}_{\mathcal{M}}$ in terms of the Jacobi bracket on $\mathcal{P}$ following the steps outlined in section \ref{sec: particle mechanics}.
First of all, we note that the vector field $X$ above is in the kernel of $\omega_{m}$ as it should be, but it satisfies $i_{X}\theta_{m}=\frac{m^{2}}{p_{0}}$.
However, the Reeb vector field $\Gamma$ is:
\be
\Gamma\,=\,\frac{\eta^{\mu\nu}p_{\mu}}{m^{2}} \frac{\partial}{\partial u^{\mu}}\,. 
\ee
The quotient space with respect to the characteristic foliation determined by $X$, or any other vector field proportional to it, is the so-called ``frozen'' phase space  \cite{komar_bergmann-frozen}.
The Jacobi structure on the algebra of functions $C^{\infty}(\mathcal{P})$ can be written in terms of $\Gamma$ and the bivector
\begin{equation}
\Lambda = \left( \delta_{\nu}^{\mu} - \frac{ p_{\nu}p_{\rho}\eta^{\rho\mu}}{m^2} \right)\frac{\partial}{\partial p_{\mu}}\wedge\frac{\partial}{\partial u^{\nu}}\,.
\end{equation}
We should stress here that while space-time coordinates are functions on $\mathcal{P}$, they would not be functions on the frozen phase-space, and therefore we would not be able to compute their Poisson bracket according to the symplectic structure defined on it. 
Then, the Jacobi bracket of two generic functions $f,g \in C^{\infty}(\mathcal{P})$ is 
\begin{equation}
\left[ f,g \right]_J = \Lambda(\dd f, \dd g) + f \mathrm{L}_{\Gamma}g - g\mathrm{L}_{\Gamma}f\,,
\end{equation}
from which one derives that the position functions $u^{\mu}$ on $\mathcal{P}$ do not commute
\begin{equation}
\left[ u^{\mu} , u^{\nu} \right]_J = \frac{u^{\mu}p^{\nu} - u^{\nu}p^{\mu} }{m^2}\,.
\end{equation}
Recall that the cotangent lift of the canonical action of the Poincar\'{e} group on $E$ leaves $\mathcal{P}$ invariant, and the Jacobi bracket $\left[ \cdot,\cdot \right]_J$ is itself invariant because both $\Lambda$ and $\Gamma$ are invariant.
The functions $J_{\mu\nu} = u_{\mu}p_{\nu} - u_{\nu}p_{\mu}$ and $p_{\mu}$ close on the Lie algebra of the Poincar\'{e} algebra, and the associated Jacobian vector fields are 
\begin{eqnarray}
& X_{\mu \nu} = u_{\mu}\frac{\partial}{u^{\nu}}- u_{\nu}\frac{\partial}{u^{\mu}} + p_{\mu}\frac{\partial}{p^{\nu}} - p_{\nu}\frac{\partial}{p^{\mu}} \\
& X_{\mu} = \frac{\partial }{\partial u^{\mu}}\,,
\end{eqnarray}
which coincide with the restriction to $\mathcal{P}$ of the infinitesimal generators of the canonical action of the Poincar\'{e} group on $\mathbf{T}^*E$. This means that the same vector fields may be thought of either as the canonical ones restricted to the ``mass-shell'', or as the Jacobi vector fields associated with ``momenta'' and ``angular-momenta'' functions directly on $\mathcal{P}$.
This means that there is a realization of the Poincar\'{e} group in terms of Jacobi vector fields.
We should notice that while the Lie bracket is well defined for vector fields, for functions we only have a Jacobi bracket, the Poisson bracket from $T^{*}E$ does not ``restrict'' to functions on the mass-shell.

A possible choice of Darboux coordinates for $\mathcal{P}$, say $(Q^j, P_j,T)$ with $j=1,2,3$, contains the so-called Newton-Wigner positions and the momenta 
\begin{equation}
Q^{j}= u^j + \frac{p^j}{\sqrt{p_jp^j + m^2}}x^0,\quad P_j = p_j
\end{equation}  
which are ``constants of the motion'' in   $\mathcal{P}\subset\mathbf{T}^{*}E$ with respect to the vector field $X$, and the dynamical time function
\be
T=p_{\mu}u^{\mu}
\ee
satsifying $\mathcal{L}_{\Gamma}T=1$ on the mass-shell. 
Clearly, if $T$ is taken to be the bundle projection, $T\colon\mathcal{P}\ra \mathcal{M}$, then $\mathcal{P}$ can be again represented as $\mathbf{T}^{*}\mathcal{Q}\times\mathbb{R}$ by means of the Darboux coordinates $(Q^j, P_j,T)$, and the resulting Hamiltonian system  is ``frozen'' in the sense that there is no evolution of the $Q^{j}$'s and the $P^{j}$'s.
This is in sharp contrast with the description {\itshape \`{a} la} Landau. 

From the above discussion it emerges that, since the notion of simultaneity depends on the choice of a reference system, the bundle structure $t\colon E\ra\mathcal{M}$ does not possess an invariant character with respect to the action of the Poincar\'{e} group. After the choice of a time function $\tau$ on $\mathcal{P}$, either kinematical or dynamical, the solutions of the variational problem associated with the action functional $S$ are integral curves of a vector field $X$. Since the bundle structure is not Poincar\'{e} invariant, the canonical action of the Poincar\'{e} group on the mass-shell $\mathcal{P}$ does not preserve the space of solutions of the Euler-Lagrange equations, which are sections of this bundle. Therefore, when we consider a one-parameter family of sections, providing a foliation of the total space $\mathcal{P}$ with respect to the projection onto $\mathcal{N}_H$, the canonical action of the Poincar\'{e} group will not preserve it and the Poisson bracket constructed out of it will not be invariant with respect to this action. Moreover the space of functions on every leaf of the foliation will only contain constants of the motion of the vector field $X$, and the physical space-time positions $u^{\mu}$ are not. On the other hand the Jacobi bracket on $C^{\infty}(\mathcal{P})$ is invariant with respect to the canonical action of the Poincar\'{e} group and it is defined also for the coordinate functions $u^{\mu}$, which are not commuting with respect to this bracket.

In order to avoid the choice of a reference system, one can work in a different setting. Indeed, without introducing a time function on $\mathcal{P}$, one can consider embeddings $\chi\colon\mathcal{M}=\mathbb{R}\mapsto\mathcal{P}$ of the real line $\mathcal{M}$ into the mass-shell, which are transversal to the fibres of the bundle $\pi_{\mathcal{P}}\colon\mathcal{P}\rightarrow E$, i.e., the map $\pi_{\mathcal{P}}\circ\chi\colon \mathcal{M}\mapsto E$ is still an embedding. The space of fields $\mathscr{F}_{\mathcal{P}}$ becomes the space of all these embeddings and a variation can be identified with a vector field along $\chi$ which, however, does not need to be vertical anymore, with respect to a possible bundle-projection onto $\mathbb{R}$. Therefore, when we apply the Schwinger-Weiss action principle we get:
\begin{equation}
\chi^*(i_{\tilde{U}}\dd \theta_m) = \delta u^{\mu}\dot{p}_{\mu} + \delta p_{\mu}\dot{u}^{\mu} = 0
\end{equation}
where $\tilde{U}=\delta u^{\mu}\frac{\partial }{\partial u^{\mu}} + \delta p_{\mu}\frac{\partial}{\partial p_{\mu}}$ satisfies 
$$\eta^{\mu \nu} \delta p_{\mu}p_{\nu}=0\,.$$ Therefore, one gets the following Euler-Lagrange equations:
\begin{equation}
\dot{p}_{\mu}=0,\quad \left( \delta^{\mu}_{\nu} - \frac{p^{\mu}p_{\nu}}{m^2} \right)\dot{x}^{\nu} = 0\,,
\end{equation}
and this set of equations is clearly invariant under reparametrization of the embedded manifold.
Therefore, the solutions are equivalence classes of curves with respect to the action of the reparametrization. Due to the invariance of the 1-form $\theta_m$, this system of equations is invariant under the canonical action of the Poincar\'{e} group on $\mathcal{P}$, and the space of solutions $\mathcal{EL}_{\mathcal{M}}$ is preserved, too. 
In order to find an expression of the bracket on $\mathcal{EL}_{\mathcal{M}}$, one can choose a parametrization, which amounts to fixing a gauge, coming back to the previous analysis. However, in this case the action of the Poincar\'{e} group will preserve the space of solutions, provided that the parametrization can be changed according to the imposed transformation. 
The Jacobi bracket, however, would remain always the same, due to its invariant behaviour.  

In conclusion we may summarize the situation in simple terms. Any dynamical system on the eight dimensional phase-space given by the cotangent bundle of space-time requires a manifold of Cauchy data (space of constants of the motion) which is seven dimensional. Families of integral curves may be selected by fixing one of the Cauchy data pertaining to the family, say the rest mass. In this way only six dimensional submanifolds of Cauchy data are required, and they carry a symplectic structure. Any attempt to introduce canonically conjugate pairs of Cauchy data with the requirement that half of them may possibly be identified with physical positions, while preserving invariance under the Poincar\'{e} group will fail. Thus, we are forced to use an odd dimensional carrier space if we wish to preserve the Poincar\'{e} invariance of our description. On this carrier space the only available Lie brackets among functions is a Jacobi bracket.

\section{Conclusions}

In this letter we presented an alternative point of view for the description of the covariant bracket on the space of functionals on solutions to a variational problem.
Specifically, we outlined the relevance of contact geometry and the Jacobi bracket naturally emerging in this context.
The space of solutions $\mathcal{EL}_{\mathcal{M}}$ to the variational problem  for non-relativistic Hamiltonian mechanics and for the relativistic particle  is shown to be diffeomorphic to a quotient of a suitable contact manifold (e.g.,  the extended phase space and the mass-shell submanifold).
Therefore, the algebra  of smooth functions on  $\mathcal{EL}_{\mathcal{M}}$ may be identified with the subalgebra of smooth functions on the unfolded contact manifold which is invariant with respect to vertical transformations.
This subalgebra turns out to be a Poisson subalgebra of the algebra of smooth functions on the contact manifold endowed with the Jacobi bracket.
In this way, an unfolded description of the Poisson bracket on smooth functions on $\mathcal{EL}_{\mathcal{M}}$ is given. 

In a companion letter \cite{C-DC-I-M-S-2020-02}, we provide some general considerations on how to extend this approach to fields and illustrate these ideas in the case of Klein-Gordon fields, and the Schr\"{o}dinger equation, proving that the contact framework and its associated Jacobi bracket provide a ``choice-invariant'' framework also in the  context of more general field theories.
Having in mind the passage to the quantum case, the analysis presented here seems to point out at the necessity of understanding the quantum analogue of contact manifolds and their Jacobi brackets by means of the operator-Lagrangian and Quantum Action Principle as proposed by Schwinger.


\section*{Acknowledgments}

F.D.C. and A.I. would like to thank partial support provided by the MINECO research project MTM2017-84098-P and QUITEMAD++, S2018/TCS-A4342. A.I. and G.M. acknowledge financial support from the Spanish Ministry of Economy and Competitiveness, through the Severo Ochoa Programme for Centres of Excellence in RD(SEV-2015/0554). G.M. would like to thank the support provided by the Santander/UC3M Excellence Chair Programme 2019/2020, and he is also a member of the Gruppo Nazionale di Fisica Matematica (INDAM), Italy.

%

\end{document}